# Characterizing Spam traffic and Spammers

Cynthia Dhinakaran and Jae Kwang Lee
*Department of Computer Engineering*
*Hannam University, South Korea*

Dhinaharan Nagamalai
*Wireilla Net Solutions Inc,*
*Chennai, India,*

## Abstract

*There is a tremendous increase in spam traffic these days [2]. Spam messages muddle up users inbox, consume network resources, and build up DDoS attacks, spread worms and viruses. Our goal is to present a definite figure about the characteristics of spam and spammers. Since spammers change their mode of operation to counter anti spam technology, continues evaluation of the characteristics of spam and spammers technology has become mandatory. These evaluations help us to enhance the existing technology to combat spam effectively. We collected 400 thousand spam mails from a spam trap set up in a corporate mail server for a period of 14 months form January 2006 to February 2007. Spammers use common techniques to spam end users regardless of corporate server and public mail server. So we believe that our spam collection is a sample of world wide spam traffic. Studying the characteristics of this sample helps us to better understand the features of spam and spammers technology. We believe that this analysis could be useful to develop more efficient anti spam techniques.*

## 1. Introduction

E-mail has emerged as an important communication source for millions of people world wide by its convenience and cost effectiveness [6]. Email provides user's low cost message to large number of people by simply clicking the send button. Email message sizes ranges from 1 kb to multiple mega bytes which is much larger than fax and other communication devices. The byproducts of email like instant messaging, chat etc., make life easier and adds more sophisticated facilities to the Internet users. According to a Radicati Group [18] study from the first quarter of 2006, there were about 1.1 billion email users worldwide. Traditionally the Internet penetration is very high in USA and Europe. But due to the recent upraise of Asian power houses like China and India, the number of email users have increased tremendously [15]. These days spam has become a serious problem to the Internet Community [8]. Spam is defined as unsolicited, unwanted mail that endangers the very existence of the e-mail system with massive and uncontrollable amounts of message [4]. Spam brings worms, viruses and unwanted data to the user's mailbox. Spammers are different from hackers. Spammers are well organized business people or organizations that want to make money. DDoS attacks, spy ware installations, worms are not negligible portion of spam traffic. According to research [5] most spam originates from USA, South Korea, and China respectively. Nearly 80% of all spam are received from mail relays [5]. Our aim is to present clear characteristics of spam and spam senders. We setup a spam trap in our mail server and collected spam for the past 14 months from January 2006 to February 2007.

We used this data for our study to characterize spam and its senders. We conducted several standard spam tests to separate spam from incoming mail traffic. The standard test includes various source filters, content filter. The various source filter tests includes Baysean filter, DNSBL, SURBL, SPF, Grey List, rDNS etc. The learning is enabled in content filters. The size of the dictionary is 50000 words. At our organization we strictly implement mail policies to avoid spam mails. The users are well instructed on how to use mail service for effective communication.

The rest of the paper is organized as follows. Section 2 discusses related work. Section 3 provides data collection of legitimate and spam mails. In section 4, we describe our classification and characteristics of spam traffic. Section 5 provides details of spammers and their technology. We conclude in section 6.

## 2. Related work

In [1] propose a novel approach to defend DDoS attack caused by spam mails. Their study reveals the effectiveness of SURBL, DNSBLs, content filters. They have presented inclusive characteristics of virus, worms and trojans accompanied spam as an attachment. Their approach is a combination of fine tuning of source filters, content filters, strictly implementing mail policies, educating user, network monitoring and logical solutions to the ongoing attack. In [3] examines the use of DNS black lists. They have examined seven popular DNSBL and found that 80% of the spam sources are listed in some DNSBL. In [4] presented a comprehensive study of clustering behavior of spammers and group based anti spam strategies. Their study exposed that the spammers has demonstrated clustering structures. They have proposed a group based anti spam frame work to block organized spammers. In [5] presented a network level behavior of spammers. They have analyzed spammers IP address ranges, modes and characteristics of botnet. Their study reveals that blacklists were remarkably ineffective at detecting spamming relays. Their study states that to trace senders the internet routing structure should be secured. In [6] presented a comprehensive study of spam and spammers technology. His study reveals that few work email accounts suffer from spam than private email. In [8] Gomez, Crsitino presented an extensive study on characteristics of spam traffic in terms of email arrival process, size distribution, the distributions of popularity and temporal locality of email recipients etc., compared with legitimate mail traffic. Their study reveals major differences between spam and non spam mails.

## 3. Data Collection

Our characterization of spam is based on 14 months collection of data over 400,000 spam from a corporate mail server. The web server provides service to 200 users with 20 group email IDs and 200 individual mail accounts. The speed of the Internet connection is 100 Mpbs for the LAN, with 20 Mbps upload and download speed (Due to security and privacy concerns we are not able to disclose the real domain name). To segregate spam from legitimate mail, we conducted a standard spam detection tests in our server. The spam mails detected by these techniques were directed to the spam trap in the mail server.

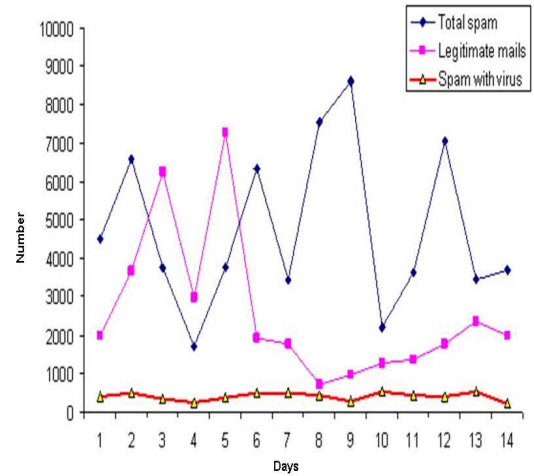

**Figure 1.** Incoming mail traffic

.Spammers do not change their tactics on day to day basis. Our study shows that spammers follow the same technology until the anti spammers find efficient way to keep them at bay. The time period ranges from 8 months to 1 year. We found that the major spammers follow the same technology from May 2006 to February 2007. The figure 1 shows the incoming mail traffic of our mail server for 2 weeks. The figure shows that the spam traffic is not related to legitimate mail traffic. The legitimate mail traffic is two way traffic induced by social network [8]. But the spam traffic is one way traffic. From this picture we can understand that the server is handling more number of spam than legitimate mail traffic.

The Figure.1 shows the number of legitimate mails, spam and spam with virus as an attachment for a period of 2 weeks from February 1 to February 15. The x axis is day and y axis is the number of spam received by the spam trap on server. Roughly the number of legitimate mails ranges from 720 to 7253 with an average rate of 906 per day. The spam mails ranges from 1701 to 8615 with an average rate of 4736 per day. The spam with viruses as an attachment ranges from 209 to 541 with an average rate of 403 per da*y*.

## 4. Spam Classification and Characteristics

We have analyzed millions of spam received in our spam trap. Mostly the spam mails are related to finance, pharmacy, business promotion, adultery services and viruses. Considerable amount of spam has virus and worms as an attachment [1].

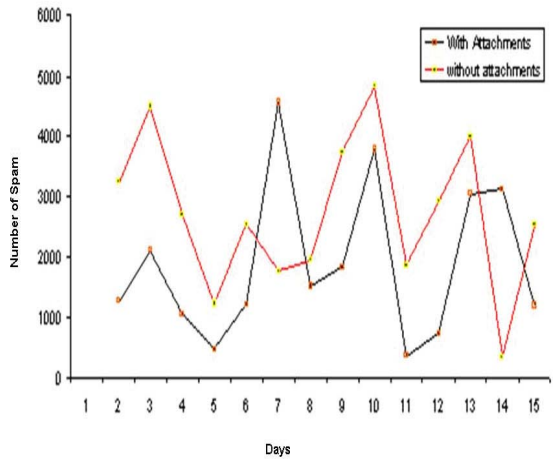

**Figure 2.** Spam with/ without attachment traffic

Based on our study the spam mails typically fall into one of two camps, like spam without attachment and spam with attachment. The Fig. 2 shows the number of spam with attachment and spam without attachment collections for a period of 2 weeks in our corporate mail server. The x axis is day and y axis is the number of spam received by our mail server. The spam with attachment mail traffic ranges from 354 to 4557 with an average rate of 1872 mails per day. The spam mails without attachment received by our spam trap ranges from 346 to 4825 with an average rate of 2722. The spam with attachment does not have any relationship with spam without attachment in the terms of traffic volume. Both are driven by different spammers with different technology.

### 4.1. Spam without attachment

Spam without attachments are mostly text messages. It can be roughly categorized as text only messages and text messages with URL or a clickable link to website. The size of these message ranges from 2 kb to 3 kb. Spam which contains only text messages are mostly related to scam. Scam related mails are not mass mails that are not well organized like other spam mails. The scam mails are sent by individual sender to limited number of receivers. Most of the sender's mail accounts are fake or don't exist. We identified some messages are sent from Africa using Japanese domains. The spammers send mails to unknown user mail accounts by assumption. The scam mails are very less in numbers when compared to other category of spam without attachment. The size of the scam mail ranges from 2 to 3 kb.

**Table 1**. Spam without attachment data collection

| Days | Plain text message | Text + link to the website | Others |
|---|---|---|---|
| 1 | 2271 | 949 | 25 |
| 2 | 1790 | 2682 | 14 |
| 3 | 207 | 2492 | 1 |
| 4 | 489 | 734 | 1 |
| 5 | 169 | 2376 | 1 |
| 6 | 253 | 1520 | 0 |
| 7 | 546 | 2167 | 5 |

Spam trap received considerable numbers of spam without attachment. These mails are related to pharmacy. These pharmacy related spam contains text messages and links to their web site. The size of these mails in this category ranges from 1 kb to 3 KB.

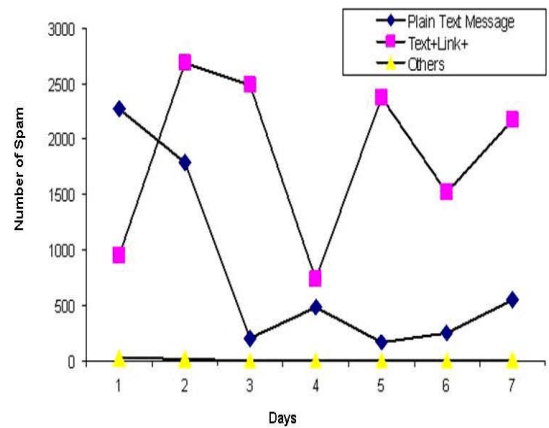

**Figure 3**. Spam without attachment collection

Figure. 3 shows the number of plain text spam mails, spam mail containing text message and a link to the targets website or a URL and different types of spam mails received by our spam trap for a period of seven days. The x axis is day and y axis is the number of spam received in the spam trap. The number of spam containing only text as a message ranges from 169 to 2271 with an average of 818 per day.

The number of spam mails with text and a clickable link or an URL ranges from 734 to 2682 with an average rate of 1846. The false positive and some spam error are notified as others in this figure. The others range from 0 to 25 with an average of 8 spam mails. We found that in a week the spam containing text and Clickable link or URL is 100% higher than spam containing only text messages. There is no relation between spam containing plain text and spam with text plus clickable link or URL in terms of traffic volume and traffic pattern.

## 4.1. Spam with attachment

In our spam collection more than 50% of the spam falls under the category of spam with attachments. This kind of spam contains a combination of pictures, text and URL or a clickable link to a web site. Mostly the attachments are image file (.gif, .jpg) and executable (.exe) files. The image files are mostly .gif format and rarely .jpg format. The size of this kind of spam ranges from 5kb to 45 kb. Spam with .exe file as an attachment is mail bombs containing viruses and worms [1]. Spam with attachments can be classified into Spam mails containing image file (.gif) and text message, spam containing image, text, URL, and Spam containing only image with clickable web link, Spam containing worms, virus, Trojans as an attachments. The size of these spam ranges from 13kb to 45 kb with attachments as Image files or executable files. We have monitored spam with attachment traffic for a period of 14 months from Jan 2006 to February 2007. The sample data is illustrated in a table 2, for a period of 7 days.

Figure. 4 shows the number of spam with Image file and text message, spam mail containing image file with clickable link or URL, spam containing text, image, URL, and other spam mails received by spam trap for a period of one week. The other spam mail category includes false positive and bulk mail once subscribed by the users. The x axis is day and y axis is the number of spam received in the spam trap. The spam containing Image file and text message ranges from 17 to 2005 with an average of 943 per day.

**Table 2**. Spam with attachment collection

| Days | Image file +text | Image +Link /URL | Image +Text +URL | Others |
|---|---|---|---|---|
| 1 | 611 | 505 | 168 | 301 |
| 2 | 1220 | 99 | 188 | 170 |
| 3 | 184 | 121 | 81 | 282 |
| 4 | 17 | 80 | 255 | 85 |
| 5 | 746 | 33 | 195 | 33 |
| 6 | 2005 | 0 | 911 | 1093 |
| 7 | 1820 | 124 | 235 | 156 |

The spam mails containing image and a clickable link or an URL ranges from 0 to 545 with an average rate of 137. The spam containing Image file, text and URL ranges from 81 to 911 with an average of 290 per day. The other spam ranges from 33 to 1093 with an average of 303 spam mails. The trap received more number of spam containing Image file with text message than other category during the week. There is no relation between these categories of spam with attachments in terms of spam traffic volume.

The size of the individual spam ranges from 8kb to 40 kb. Each spam was being sent by unique ids from unique IP addresses. The spammer used large number of IP addresses with different mail accounts. Mostly these spam were received from spoofed IP addresses with fake mail accounts and by using relays [9] [5]. The subscribed mails like news letters and commercials from companies like telecom, home appliances, are also considered as a spam by most users. These kinds of spam are easy to unsubscribe and it is negligible in the total spam percentage. The subscribed mails have pictures, text, and animation as contents with more than 50 kb as their size.

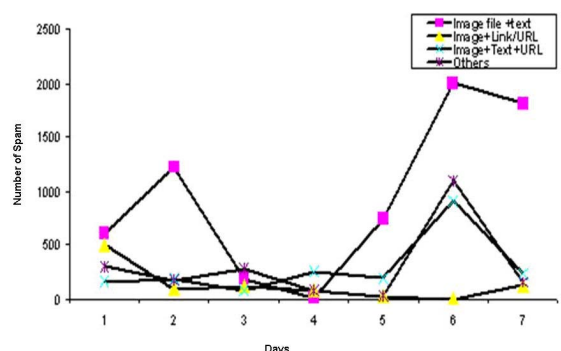

**Figure 4**. Spam with attachment collection

Spam contains text, image, URL: This category of spam contains text, Image and URL. The text is unrelated to the intended spammers business. The Image contains URL which is directly linked to the intended business organization. The size of the spam ranges from 10 kb to 30 kb. Considerable portion of these kinds of spam are responsible for phishing attacks.

Spam containing Image and text: In this category we discuss spam containing text and Image as a message. The image contains text and URL of online business. Mostly this kind of spam doesn't have clickable links. The contents of the image are relevant to the intended business organization. But the text messages following the images are randomly selected and are not related to the intended message. The text message size varies from a single line to several paragraphs. By changing the text contents and its size, the spammers try to confuse the filters. In this category we found that the majority of spam are related to pharmacy and finance. Our study shows in this category more than 50% of the spam are related to financial matters. The size, color, contents of the picture is same for a particular period of the time. The spammer's main intention is to deliver a message available inside the picture. To confuse the filters, the spammers add different size of text message. Spam containing image and URL or a clickable link: Considerable number of spam contains image file and URL or a clickable link. Since the URL or clickable link is placed inside the image, the spam can easily bypass content filters.

The fourth category of spam contains .exe file as an attachment. The spam with .exe file as an attachments are mostly virus, worms, trojon etc,. The attachment sizes ranges from 35 k to 140 kb. These mails are intended to spread virus, try to establish mail bombs to mount DDoS attack to the server and the network. Upon execution of the attachment, it will drop new files in windows folder and change the registry file, link to the attacker's website to download big programs to harm the network further. The infected machine collects email addresses from windows address book and automatically sends mails to others in the same domain [1].

# 5 Spammers Technology

We observed that the spammers use software tool to send spam with attachment. The software provides sophisticated facilities to spam end users. The characteristics of spam software are hiding the sender identity, randomly selecting text messages, identifying open relay machines, mass mailing capability, defining the spamming time and duration. We also identified that spammers manually send spam without attachment to limited number of end users. Spammers don't use spam software to send spam without attachment.

## 5.1 Spam without attachment

In case of text message with link, we have noticed 3 categories of senders. The first type of spammers are using free mail service providers like yahoo and other unfamiliar mail services. In this case it is single person sends spam to many users with multiple fake ids with the same contents. Mostly the number of receivers are 2, rarely it exceeds the limit 2, and if it does it usually does to less than 10. The destination mail accounts are normally collected from spammers or by assumption. The contents of spam include pharmacy, adult services and finance etc. The link available is not directly related to the spammer's website. The link available is a legitimate website or blog, from there it will be redirected to the spammers website. The legitimate website is used to confuse the anti-spam techniques. We have observed that spammers use Yahoo, geocities as their intermediate machine to redirect their link as shown in the link below.

- http://us.rd.yahoo.com/*http://jpsvpd.miserpoison. com/legalrx/?38947779
- http://www.geocities.com/hi_bumukybew/ redirected to http://www.enjoycasualsex.com/

Our study shows that mostly these categories of spam mail originate from US. The second category of spammers own small size web spaces with limited mail accounts. The website has unwanted data or "under construction" notice. Most spammers use European countries or Asia Pacific region ISPs as a host. Mainly contents of the spam related to Casinos and Pharmacy. Plain text massages are very small in number when compared to other kind of spam. Example for this kind of mail is the Nigerian Fraud Email, etc, [15][16]

## 5.2 Spam with attachment

The senders of spam with attachment are highly sophisticated. Mostly the spam sent are by using well designed software or program. Form our analysis we have identified that 100% of the sender's mail accounts don't exist i.e. fake mail accounts. Spammer doesn't have own website or domain. The spam shows only one receiver in the mail but that is not true. Mostly those spam arrive from a relay machine which makes it difficult to find the real sender. Spammers use spoofed ip address to hide their identity.

There are many software available to make spamming more sophisticated and easy [1]. For example Phasma Email Spoofer, Bulk Mailer, Aneima 2.0, Avalanche 3.5, Euthanasia etc. Spam software use random servers to create fake mail accounts and the spammer can fix number of mails to send and the period of the attack. It does facilitate the spammer to use relay machines. By using this software spammer can send mass mail to end users without showing all destination mail accounts. Spammer can give guidelines to the software to create a header else it will generate its own header. The spammer can attach their intended file as an attachment with random or fixed text message. There are software which can send 365 spam per minute. The mail sent by using sophisticated software pretends to be sent by Microsoft outlook express [12]. Since the messages contain only HTML content, these messages were not sent by using Microsoft outlook express. But the spam message contain virus, worms, trojon as an attachment doesn't show that they were sent by Microsoft outlook express. Regardless of spam type whether with or without attachments mostly spam sent with the help of software indicates that which are sent by Microsoft outlook express.

## 5. Conclusions

From our study we conclude that spam can be classified into two wide categories. The first category is spam with attachment and the second category is spam without attachments. Spam without attachment are text messages and links to the intended target. Spam with attachments can be classified into 4 types such as Spam mails containing image file (.gif) and text message, spam containing image, text, URL, spam containing only image with clickable web link and spam containing worms, virus, and trojans as an attachments. Spam without attachments are small in size but spam with attachment is bigger in size. The volume of spam traffic is not related to legitimate mail traffic. From our study we tried to understand the technology used by spammers. For spam without attachment, senders use non sophisticated methods. They send spam without the help of software and use free mail service providers like Yahoo etc. But for spam with attachment, senders use sophisticated software to spam end users. The spamming software provides various facilities like multi targets, spoofing identity, using relays etc., Due to this, tracing senders is difficult.